\documentclass[aps,prl,twocolumn,groupedaddress]{revtex4}

\usepackage{graphicx}
\usepackage{calc}
\usepackage{bm}

\begin{document}


\title{Waveform sampling using an adiabatically driven electron ratchet in a two-dimensional electron system}


\author{T.~M\"uller}
\email[Corresponding~author, e-mail: ]{t.mueller@gmx.net}
\affiliation{Laboratorium f\"ur Festk\"orperphysik, Universit\"at
Duisburg-Essen, Lotharstr. 1, D-47048 Duisburg, Germany}
\author{A.~W\"urtz}
\altaffiliation[Present address: ]{ATMEL Germany GmbH,
Theresienstr. 2, D-74072 Heilbronn, Germany}
\affiliation{Laboratorium f\"ur Festk\"orperphysik, Universit\"at
Duisburg-Essen, Lotharstr. 1, D-47048 Duisburg, Germany}
\author{A.~Lorke}
\affiliation{Laboratorium f\"ur Festk\"orperphysik, Universit\"at
Duisburg-Essen, Lotharstr. 1, D-47048 Duisburg, Germany}
\author{D.~Reuter}
\affiliation{Lehrstuhl f\"ur Angewandte Festk\"orperphysik,
Ruhr-Universit\"at Bochum, Universit\"atsstr. 150, D-44780 Bochum,
Germany}
\author{A.~D.~Wieck}
\affiliation{Lehrstuhl f\"ur Angewandte Festk\"orperphysik,
Ruhr-Universit\"at Bochum, Universit\"atsstr. 150, D-44780 Bochum,
Germany}
\affiliation{}


\date{\today}

\begin{abstract}
We utilize a time-periodic ratchet-like potential modulation
imposed onto a two-dimensional electron system inside a
GaAs/Al$_x$Ga$_{1-x}$As heterostructure to evoke a net dc pumping
current. The modulation is induced by two sets of interdigitated
gates, interlacing off center, which can be independently
addressed. When the transducers are driven by two identical but
phase-shifted ac signals, a lateral dc pumping current
$I(\varphi)$ results, which strongly depends on both, the phase
shift $\varphi$ and the waveform $V(t)$ of the imposed gate
voltages. We find that for different periodic signals, the phase
dependence $I(\varphi)$ closely resembles $V(t)$. A simple linear
model of adiabatic pumping in two-dimensional electron systems is
presented, which reproduces well our experimental findings.
\end{abstract}


\maketitle Most studies on electronic transport in nanostructures
are based on transport experiments, where a current is driven
through the system by biasing at least one contact using an
external electrical potential. Up to now, only few experiments
were reported, where the current is generated by a temporal and
spatial periodic potential modulation in an externally unbiased
system, as proposed in refs. \cite{theorie}. These systems closely
resemble ratchets and molecular motors,
\cite{feynman1966,ratchets} which attracted much attention in
recent years, due to their extraordinary high efficiency to
transform heat into directed motion. Experimental realizations of
these so-called electron pumps range from arrays of
$\mathrm{Al}/\mathrm{Al}_2\mathrm{O}_3/\mathrm{Al}$ tunneling
junctions,
\cite{delsing1989,geerligs1990,pothier1992,martinis1994,keller1999,lotkhov2001}
open and closed artificial semiconductor quantum dot structures,
\cite{kouwenhoven1991,switkes1999} electrons captured in the
potential landscape of moving surface acoustic waves \cite{saw} to
semiconductor devices like charge coupled devices (CCD), nowadays
routinely used for digital photographic imaging. \cite{ccd} While
most experiments utilizing parametric pumping are carried out in
the Coulomb blockade (CB) regime,
\cite{delsing1989,geerligs1990,pothier1992,martinis1994,keller1999,lotkhov2001,kouwenhoven1991}
only few experiments so far are performed in the open regime.
\cite{switkes1999,hoehberger2001} CB systems already serve as
metrological standards for both, the electrical current
\cite{delsing1989,geerligs1990,pothier1992,martinis1994,lotkhov2001,kouwenhoven1991}
and the capacitance, \cite{keller1999} reaching relative errors of
$10^{-11}$ \cite{lotkhov2001} and $10^{-6}$, \cite{keller1999}
respectively.

In this letter we present electronic transport experiments in a
laterally confined strip of a two-dimensional electron system
(2DES). By imposing a spatial and temporal periodic potential
perturbation, \cite{hoehberger2001} using interdigitated gates,
similar to a CCD device, a dynamic ratchet-like potential
landscape is created which drives a net lateral dc pumping
current.
\begin{figure}
\includegraphics{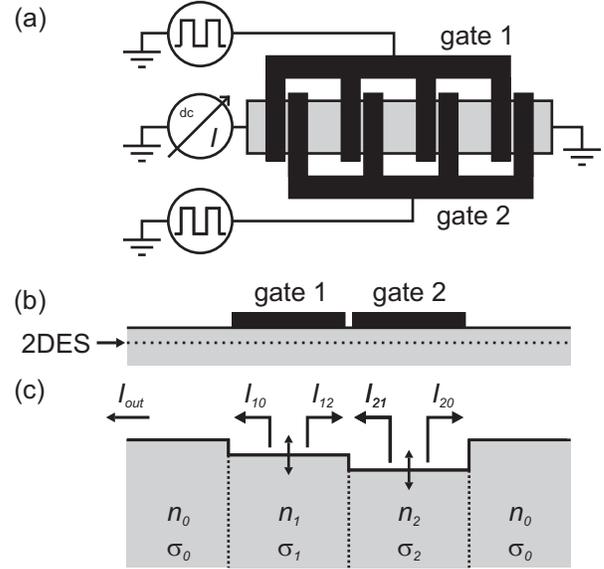}
\caption{(a) Schematic top view of the sample. Two sets of
metallic interdigitated gates (black) are evaporated on top of a
mesa stripe (gray). By applying time-periodic signals to the
gates, a dc source-drain current is induced. (b) Cross sectional
view of the sample. (c) Pumping mechanism, for details please
refer to the text.\label{fig1}} \end{figure} The sample presented
here is fabricated from a molecular beam epitaxy grown
GaAs/Al$_x$Ga$_{1-x}$As heterostructure, containing a 2DES
$54.8\:\mathrm{nm}$ below the surface. The electron density $n$
and mobility $\mu$ at $4.2\:\mathrm{K}$ are $4.13 \times
10^{15}\:\mathrm{m}^{-2}$ and $94.5\:\mathrm{m}^2/\mathrm{Vs}$,
respectively. Samples from other heterostructures (not presented
here) yield similar results. A Hall bar-shaped mesa is fabricated
using optical lithography, etching, contact metallization and
annealing. On top, two sets of interdigitated gate transducers
(IGTs) are defined by electron beam lithography and subsequent
metallization. Each set consists of 75 gate stripes, (each
$250\:\mathrm{nm}$ wide) interlacing off center by two thirds of
the lattice period of $1\:\mu\mathrm{m}$. A schematic top view of
the sample is shown in Fig. \ref{fig1}(a). Measurements are
performed inside a ${}^3\mathrm{He}$ cryostat at a base
temperature of $300\:\mathrm{mK}$. \cite{temp} The dc source-drain
pumping current is measured using a transimpedance-converter,
whose output signal is filtered by an eight-pole
$48\:\mathrm{dB}/\mathrm{oct}$ low-pass Bessel-filter, set to a
corner frequency of $10\:\mathrm{Hz}$. The same waveform -- with a
well-defined phase difference $\varphi$ -- is applied to the two
sets of gates, using two frequency-locked arbitrary waveform
generators. The signals applied to gates 1 and 2 can be expressed
as $V_1(t)=-V_0/2+V_0 W(2 \pi f t)$ and $V_2(t)=-V_0/2+V_0 W(2 \pi
f t - \varphi)$, respectively, where $W(t,\varphi)$ represents one
of the normalized \cite{amplitude} waveforms -- sine, rectangle or
triangle -- $f$ the frequency, $t$ the time and $V_0$ the
amplitude. In order to eliminate parasitic currents, which, i. e.,
result from thermal voltages or the rectification of capacitive
currents inside the non-ideal AuGe-contacts, we measure the output
current $I_{\mathrm{out}}(\varphi)$ in the interval
$[0^\circ;360^\circ]$ and define the pumping current
$I(\varphi)=0.5 \cdot
\left(I_{\mathrm{out}}(\varphi)-I_{\mathrm{out}}(360^\circ-\varphi)\right)$
for $\varphi \in [0^\circ;180^\circ]$.

\begin{figure}
\includegraphics{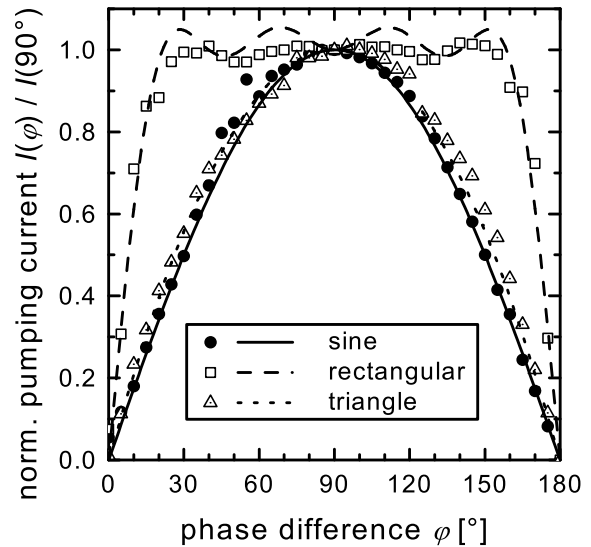}
\caption{Experimental phase dependence of the normalized pumping
current $I(\varphi)/I(90^\circ)$ for different waveforms (symbols)
at a frequency of $90\:\mathrm{kHz}$ and an excitation voltage of
$0.35\:\mathrm{V_{pp}}$. The lines represent the results of the
simulation, as described in the text.\label{fig2}}
\end{figure}

Figure \ref{fig2} shows experimental, normalized
$I(\varphi)/I(90^\circ)$-traces for different waveforms at a
frequency of $90\:\mathrm{kHz}$ and an amplitude
$V_0=0.35\:\mathrm{V_{pp}}$. The symbols represent the
experimental results. The fact, that the pumping current as a
function of the phase difference, especially for the sine and
rectangular signals, closely resembles the waveform of the pumping
signal is somewhat surprising and will be analyzed in more detail
below. Since the phase can be changed arbitrarily slow, this
observation may open possibilities for novel wave sampling
devices, even for very high frequencies.

For a deeper understanding whether our experimental results are
coincidental for the present device or are a general feature of
adiabatically driven ratchets in 2DESs, we develop a simple
hydromechanical model of adiabatic electron pumping in the
artificial superlattice. It is based on a single unit cell (one
period of the superlattice), as shown in Figs. \ref{fig1}(b) and
\ref{fig1}(c). As, in the present experiment, we use temporal
periods which are much longer than all relevant relaxation times
of the electron system, we can use an adiabatic approximation. For
a treatment of the high frequency limit, see ref.
\cite{dittrich2003}. The unit cell is divided in three regions, as
can be seen in Fig. \ref{fig1}(c). Due to capacitive coupling
between the gate electrodes and the 2DES, we assume a linear
relationship between the applied gate voltage $V_i$ and the
electron density $n_i$ inside the respective region $i$. If we
restrict the discussion to small gate voltages only ($V_D< V_i
<0$, where $V_D=-0.45\:\mathrm{V}$ is the voltage at which the
2DES is fully depleted), we can assume the conductivity $\sigma_i$
to be only depending on $n_i$, via $\sigma_i=n_i\mu e$, neglecting
small changes of the mobility $\mu$. By changing $V_1$ the
electrons have to move from (to) neighboring regions 0 and 2 into
(out of) region 1. The ratio
$I_{10}/I_{12}=\sigma_{0}/\sigma_{2}=n_0/n_2$ of electrons moving
from (to) area 0 and 2, respectively, is determined by Kirchhoff's
laws. Using this result, one can derive, i. e., the current from
region 1 into region 2:
\begin{equation}
I_{12}= \frac {\sigma_2}{\sigma_0+\sigma_2} \frac {\partial \left(
e n_1 \right)}{\partial t} = e \frac {n_2}{n_0+n_2} \frac
{\partial n_1}{\partial t}.
\end{equation}
This results in an output current $I_{\mathrm{out}}$ of the unit
cell of
\begin{eqnarray}
I_{\mathrm{out}}&=&I_{10}+I_{21} \nonumber
\\&=&e \left[ \frac {n_0}{n_0+n_2} \frac {\partial
n_1}{\partial t}+\frac {n_1}{n_1+n_0} \frac {\partial
n_2}{\partial t}\right]
\end{eqnarray}
By inserting Fourier series expansions of the respective
time-dependent electron-densities $n_i(t)$ (as defined by the
waveforms applied to the IGTs), one can evaluate the dc pumping
current $I$ of the device, by integrating the output current
$I_{\mathrm{out}}$ over one temporal period of the input waveform.
For the sine waveform, the result is plotted in Fig. \ref{fig2} as
a solid line which reproduces the experimental findings very well.

In order to model more complex waveforms, we have to consider that
in any real sample, a RC cutoff frequency $f_{\mathrm{cutoff}}$
will be present, which suppresses the higher Fourier components of
the modulation signal. To account for this experimentally
determined cutoff frequency of $\approx 300\:\mathrm{kHz}$, we
have weighted the Fourier components by a 3-term-Blackman-Harris
$\nu_{\mathrm{BH}}(f)$ distribution. \cite{blackman-harris} If we
assume $\nu_{\mathrm{BH}}(f_{\mathrm{cutoff}})=e^{-1}$, we only
have to include the first seven Fourier components into our
calculation. The results are plotted as dashed and dotted lines
into Fig. \ref{fig2} for the rectangular and triangular waveforms,
respectively. They match remarkably close the experimental data,
although effects like the gate voltage dependent mobilities
$\mu_i$ are neglected.
\begin{figure}
\includegraphics{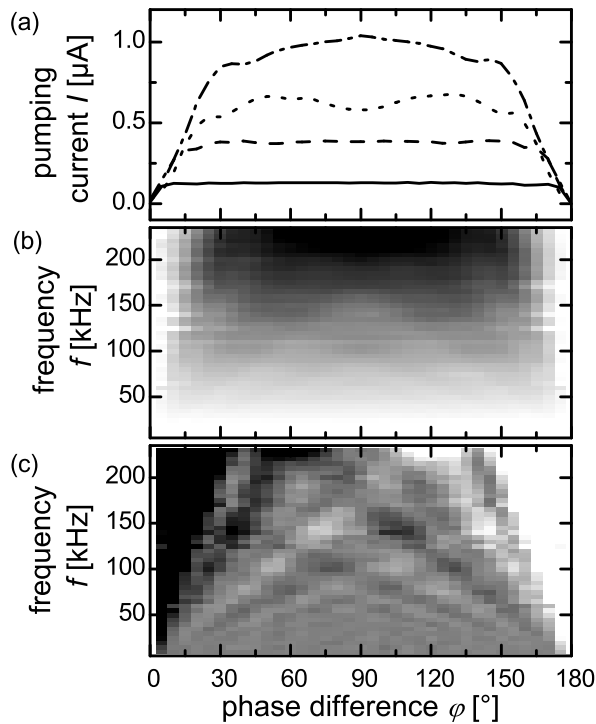}
\caption{Rectangular waveform signals for an excitation voltage of
$0.35\:\mathrm{V_{pp}}$ at different frequencies. (a)
$I(\varphi)$-traces (solid line: $30\:\mathrm{kHz}$, dashed line:
$90\:\mathrm{kHz}$, dotted line: $150\:\mathrm{kHz}$ and
dashed-dotted line: $210\:\mathrm{kHz}$), (b) gray scale plot of
$I$ and (c) gray scale plot of the $\partial I/\partial\varphi$.
Darker gray tones indicate (b) higher currents and (c) higher
derivatives $\partial I/\partial\varphi$, respectively.}
\label{fig3}
\end{figure}

The rectangular waveform pumping current is plotted for different
frequencies in Fig. \ref{fig3}(a). It can be clearly seen, that
 the pumping current increases linearly with the frequency of the
gate signal. \cite{hoehberger2001} For low frequencies, i.e.
$10\:\mathrm{kHz}$, the resulting output waveform nearly perfectly
resembles the imposed rectangular waveform. Additional
oscillations in the signal can be observed, especially for higher
frequencies. They are a direct consequence of the RC constant of
the system, which results in higher harmonics being more
efficiently attenuated at higher operating frequencies. With
increasing $f$ the traces become more and more `softened'.
Ultimately for $f>f_{\mathrm{cutoff}}$ (not shown here) the
current is determined only by the first harmonic, resulting in a
nearly sinusoidal output current. To illustrate that the
oscillations in the pumping efficiency indeed originate from the
effective filtering of the high frequency contributions, a gray
scale plot of $I(\varphi, f)$ is presented in \ref{fig3}(b). In
order to emphasize the fine-structure, the derivative $\partial I
/
\partial \varphi$ is numerically calculated and plotted against $\varphi$ and $f$ in
fig. \ref{fig3}(c). Apart from the flanks (dark area on the left
and bright area on the right hand side of the figure) the
oscillatory features extend diagonally from the lower corners of
the plot, corresponding to the harmonic components of the
rectangular waveform, as predicted by the model.

This damping of the higher harmonics also explains, why the
triangular waveform is not reproduced very well. For the
triangular waveform the unattenuated Fourier coefficients scale
like $(2n-1)^{-2}$, while for the rectangle they scale like
$(2n-1)^{-1}$, with $n=1,2,3,....\:$. So the higher Fourier
coefficients of the triangular waveform are per-se much weaker
than the respective coefficients for the rectangular waveform,
resulting in an $I(\varphi)$ characteristics, which is mainly
dominated by lower harmonics, in particular the first one.

We have presented a method to sample an arbitrary periodic
waveform by using interdigitated gates on top of a 2DES. The high
frequency operating limit is intrinsically determined by the RC
constant of the system, as this leads to the attenuation of higher
harmonics. The results were explained within a simple
hydromechanical model of adiabatic electron pumping. The described
system may serve as the foundation for future signal sampling
devices in solid state electronics. Following our
proof-of-principle, better device parameters will lead to a higher
cutoff frequency, eventually allowing a very high frequency
operation of the device.


\begin{thebibliography}{99}
\bibitem{theorie} D. J. Thouless, Phys. Rev. Lett. {\bf 27}, 6083
(1983), P. W. Brouwer, Phys. Rev. B {\bf 58}, R10135 (1998), B. L.
Altshuler and L. I. Glazman, Science {\bf 283}, 1864 (1999), and
references therein.
\bibitem{ratchets} For a review, see i. e. P. Reimann, Phys. Rep. {\bf 361}, 57 (2002)
and the special issue of Appl. Phys. A {\bf 75}, (2002).
\bibitem{feynman1966} R. P. Feynman, R. B. Leighton, and M. Sands,
\emph{The Feynman Lectures in Physics} (Addison-Wesley, Reading,
MA, 1966).
\bibitem{pothier1992} H. Pothier \emph{et al.}, Europhys. Lett.
{\bf 17}, 249 (1992).
\bibitem{lotkhov2001} S. V. Lotkhov \emph{et al.},
Appl. Phys. Lett. {\bf 78}, 946 (2001).
\bibitem{delsing1989} P. Delsing \emph{et al.}, Phys. Rev. Lett.
{\bf 63}, 1861 (1989).
\bibitem{geerligs1990} L. J. Gerlings \emph{et al.}, Phys. Rev.
Lett. {\bf 64}, 2691 (1990).
\bibitem{martinis1994} J. M. Martinis, M. Nahum, and H. D. Jensen,
Phys. Rev. Lett. {\bf 72}, 904 (1994).
\bibitem{keller1999} M. W. Keller \emph{et al.}, Science {\bf 285}, 1706
(1999).
\bibitem{kouwenhoven1991} L. P. Kouwenhoven \emph{et al.}, Phys.
Rev. Lett. {\bf 67}, 1626 (1991).
\bibitem{switkes1999} M. Switkes \emph{et al.}, Science {\bf 283}, 1905
(1999).
\bibitem{saw} V. I. Talyanksii \emph{et al.}, Phys. Rev. B {\bf 56}, 15180
(1997) and M. Rotter \emph{et al.}, Phys. Rev. Lett. {\bf 82},
2171 (1999).
\bibitem{ccd} For a review, see i. e., W. G. Ong, \emph{Modern MOS
Technology: Processes, Devices and Design} (McGraw-Hill,
Singapore, 1987).
\bibitem{hoehberger2001} E. M. H\"{o}hberger \emph{et al.}, Appl. Phys. Lett. {\bf 78}, 2905 (2001).
\bibitem{temp} For the pumping principle the temperature is of little relevance and similar results have been found up to a temperature of $77\:\mathrm{K}$, where gate leakage currents become detrimental to the experiment.
\bibitem{amplitude} Normalized in this context implies a
peak-to-peak value of 1 for each given waveform.
\bibitem{dittrich2003} T. Dittrich, M. Guti\'{e}rrez, and G.
Sinuco, Physica A {\bf 327}, 145 (2003).
\bibitem{blackman-harris} H. G\"{u}nzler and H. U. Gremlich,
\emph{IR-Spectroscopy} (Wiley-VCH, Weinheim, 2002).

\end{thebibliography}
\end{document}